\documentclass[fleqn,usenatbib]{mnras}

\usepackage{newtxtext,newtxmath}

\usepackage[T1]{fontenc}
\usepackage{ae,aecompl}

\usepackage{graphicx}	
\usepackage{amsmath}	
\usepackage{amssymb}	

\title[X-ray study of Eye of Horus]{X-ray study of the double source plane gravitational lens system Eye of Horus observed with XMM-Newton}

\author[K. Tanaka et al.]{
Keigo Tanaka$^{1}$\thanks{E-mail: tanaka@astro.s.kanazawa-u.ac.jp (Kanazawa University)},1
Ayumi Tsuji$^{1}$,
Hiroki Akamatsu$^{2}$,
J. H. H. Chan$^{3}$,
Jean Coupon$^{4}$,\newauthor
Eiichi Egami$^{5}$,
Francois Finet$^{6}$,
Ryuichi Fujimoto$^{7}$,
Yuto Ichinohe$^{8}$,
Anton T. Jaelani$^{9}$,\newauthor
Chien-Hsiu Lee$^{6}$,
Ikuyuki Mitsuishi$^{10}$,
Anupreeta More$^{11}$,
Surhud More$^{12}$,\newauthor
Masamune Oguri$^{12,13,14}$,
Nobuhiro Okabe$^{15}$,
Naomi Ota$^{16}$,
Cristian E. Rusu$^{6}$,\newauthor
Alessandro Sonnenfeld$^{12}$,
Masayuki Tanaka$^{17}$,
Shutaro Ueda$^{18}$, and Kenneth C. Wong$^{12,17}$
\\
\\
$^{1}$Graduate School of Natural Science \& Technology, Kanazawa University, Kakuma-machi, Kanazawa, Ishikawa 920-1192, Japan\\
$^{2}$SRON Netherlands Institute for Space Research, Sorbonnelaan 2, 3584 CA Utrecht, the Netherlands\\
$^{3}$Institute of Physics, Laboratory of Astrophysics, Ecole Polytechnique F\'{e}d\'{e}rale de Lausanne (EPFL), Observatoire de Sauverny, 1290 Versoix, Switzerland\\
$^{4}$Department of Astronomy, University of Geneva, ch. d\'{E}cogia 16, 1290 Versoix, Switzerland\\
$^{5}$Steward Observatory, University of Arizona, 933 North Cherry Avenue, Tucson, AZ 85721, USA\\
$^{6}$Subaru Telescope, NAOJ, 650 N Aohoku Pl, Hilo, HI 96720, USA\\
$^{7}$Faculty of Mathematics and Physics, Kanazawa University, Kakuma-machi, Kanazawa, Ishikawa 920-1192, Japan\\
$^{8}$Department of Physics, Rikkyo University, Nishi Ikebukuro 3-3-4-1, Toshimaku, Tokyo, Japan\\
$^{9}$Department of Physics, Kindai University, Higashi-Osaka, Osaka 577-8502, Japan\\
$^{10}$Department of Physics, Nagoya University, Furo-cho, Chikusa-ku, Nagoya, Aichi 464-8602, Japan\\
$^{11}$Inter-University Centre for Astronomy and Astrophysics, Post Bag 4, Ganeshkhind, Pune 411 007, India\\
$^{12}$Kavli IPMU (WPI), UTIAS, The University of Tokyo, Kashiwa, Chiba 277-8583, Japan\\
$^{13}$Research Center for the Early Universe, University of Tokyo, 7-3-1 Hongo, Bunkyo-ku, Tokyo 113-0033, Japan\\
$^{14}$Department of Physics, University of Tokyo, 7-3-1 Hongo, Bunkyo-ku, Tokyo 113-0033, Japan\\
$^{15}$Department of Physical Science, Hiroshima University, 1-3-1 Kagamiyama, Higashi-Hiroshima, Hiroshima 739-8526, Japan\\
$^{16}$Department of Physics, Nara Women's University, Kitauoyanishi-machi, Nara, 630-8506, Japan\\
$^{17}$National Astronomical Observatory of Japan, Mitaka, Tokyo 181-8588, Japan\\
$^{18}$Academia Sinica Institute of Astronomy and Astrophysics (ASIAA), No. 1, Section 4, Roosevelt Road, Taipei 10617, Taiwan\\
}

\date{Accepted XXX. Received YYY; in original form ZZZ}

\pubyear{2019}

\begin{document}
\label{firstpage}
\pagerange{\pageref{firstpage}--\pageref{lastpage}}
\maketitle

\begin{abstract}

A double source plane (DSP) system is a precious probe for the density profile of distant galaxies and cosmological parameters. However, these measurements could be affected by the surrounding environment of the lens galaxy. Thus, it is important to evaluate the cluster-scale mass for detailed mass modeling. We observed the {\it Eye of Horus}, a DSP system discovered by the Subaru HSC--SSP, with XMM--Newton. We detected two X-ray extended emissions, originating from two clusters, one centered at the {\it Eye of Horus}, and the other located $\sim100$ arcsec northeast to the {\it Eye of Horus}. We determined the dynamical mass assuming hydrostatic equilibrium, and evaluated their contributions to the lens mass interior of the Einstein radius.
The contribution of the former cluster is 
$1.1^{+1.2}_{-0.5}\times10^{12}~M_{\odot}$, which is $21-76\%$ of the total mass within the Einstein radius. The discrepancy is likely due to the complex gravitational structure along the line of sight. On the other hand, the contribution of the latter cluster is only $\sim2\%$ on the {\it Eye of Horus}. Therefore, the influence associated with this cluster can be ignored.

\end{abstract}




\section{Introduction}

\indent The {\it Eye of Horus} is a strong gravitational lens object \citep[]{MTanaka} that was discovered by the Subaru HSC--SSP survey \citep[HSC--SSP;][]{HSC1stDR,HSC1styrOverview,Miyazaki18HSC,Komiyama18HSC,Kawanomoto18HSC,Furusawa18HSC,Bosch18HSC,Haung18HSC,Coupon18HSC}.  It is known as a precious double source plane (DSP) object.
The lens galaxy of the {\it Eye of Horus} is a massive (stellar mass $\sim7\times10^{11}M_{\odot}$) early-type galaxy which is located at (RA, DEC)$=$($14^{\rm h}24^{\rm m}49^{\rm s}.0$, $-00\degr 53\arcmin 21\farcs 65$) (J2000), and the redshift of the lens galaxy is $z=0.795$. Since there are two background galaxies behind the lens galaxy whose redshifts are $z=1.302$ and $1.988$, respectively, an Einstein ring and an arc corresponding to these background galaxies are projected near the Einstein radii.

\indent DSP system is a precious probe for the gravitational structure of lens galaxies and cosmological parameters. In a distant galaxy at $z\sim0.8$, it is usually difficult to determine the gravitational structure inside the galaxy using only the stellar distribution. However, when a lens galaxy has a double source lens image, the gravitational structure at 10--100 kpc can be determined by combining lensing and dynamics, even if it is located at $z\sim0.8$ \citep[e.g.,][]{Sonnenfeld2012}. 
The matter density parameter $\Omega_{\rm M}$, the equation of state parameter $\omega$, and the Hubble parameter by cosmic microwave background (CMB) measurements \citep[]{WMAP7year} degenerate.
On the other hand, measurements of the ratio of two Einstein radii in a DSP object enables us to break the degeneracy between the Hubble parameter and the other two parameters \citep[][]{Collett2012,Collett2014,Linder2016}.

\indent The above constraints are obtained with high accuracy if the lens galaxy is an isolated system. However, \citet{Keeton2004} suggested that if it lies in a group or a cluster of galaxies, the lens image would be affected by the environment surrounding the lens galaxy, depending on the distance between lens galaxy and the group/cluster center. The lens galaxy of the {\it Eye of Horus} corresponds to the brightest cluster galaxy (BCG) of HSC J142449-005322 (hereafter the main cluster) in the CAMIRA cluster catalog \citep{CAMIRA_list}.
The richness of this cluster is $N_{\rm gal}\sim34$, and its redshift is $z\sim0.801$.
However, BCGs identified by optical data are known to often deviate from the cluster center up to several hundred kpc \citep{CAMIRA_list}. Therefore, we should determine the cluster center position accurately using X-ray data. Furthermore, \citet{CAMIRA_list} suggested that there is another cluster (HSC J142456-005157, hereafter the north-east (NE) cluster) located at (RA, DEC)$=$($14^{\rm h}24^{\rm m}56^{\rm s}.4$, $-00\degr 51\arcmin 57\farcs 39$) (J2000). The richness is $N_{\rm gal}\sim 37$, and its redshift is $z\sim 0.768$. These two clusters are separated only by $\sim 2'$, and photometric data suggests that the difference of the redshifts is only $\Delta z\sim0.03$, so the gravitational potential of the NE cluster may affect the lens image. If these two clusters are a merger, the lens model of the {\it Eye of Horus} may be complicated.

\indent X-ray observations provide us with crucial information on the cluster-scale environment. 
We observed X-ray emission surrounding the {\it Eye of Horus} with {\it XMM--Newton} and report the results in this paper. Two extended emissions were detected at the position of these two clusters (\S\ref{sec:obs}). We fitted the X-ray image to determine the center position and the extension of the intracluster medium (ICM) of each cluster (\S\ref{sec:img}). We fitted their spectra to determine the temperature of the ICM (\S\ref{sec:spec}). We calculated the cluster mass assuming hydrostatic equilibrium (hereafter hydrostatic mass), based on the parameters obtained by image fitting and spectral fitting, and we evaluated the effect of the cluster-scale dynamical mass on the lens potential (\S\ref{sec:conc}).

\indent Throughout this paper, we adopt a Hubble constant of $H_0=70$\,km\,s$^{-1}$\,Mpc$^{-1}$, and cosmological density parameters of $\Omega_{\rm M}=0.27$ and $\Omega_{\Lambda}=0.73$. $1''$ corresponds to 7.64 kpc at the main cluster and 7.52 kpc at the NE cluster, respectively. The solar abundance table by \citet{lodders2009} is used. All error ranges are 68\% confidence intervals unless otherwise stated.


\section{Observation and data reduction}
\label{sec:obs}
\begin{figure}
\includegraphics[width=\columnwidth]{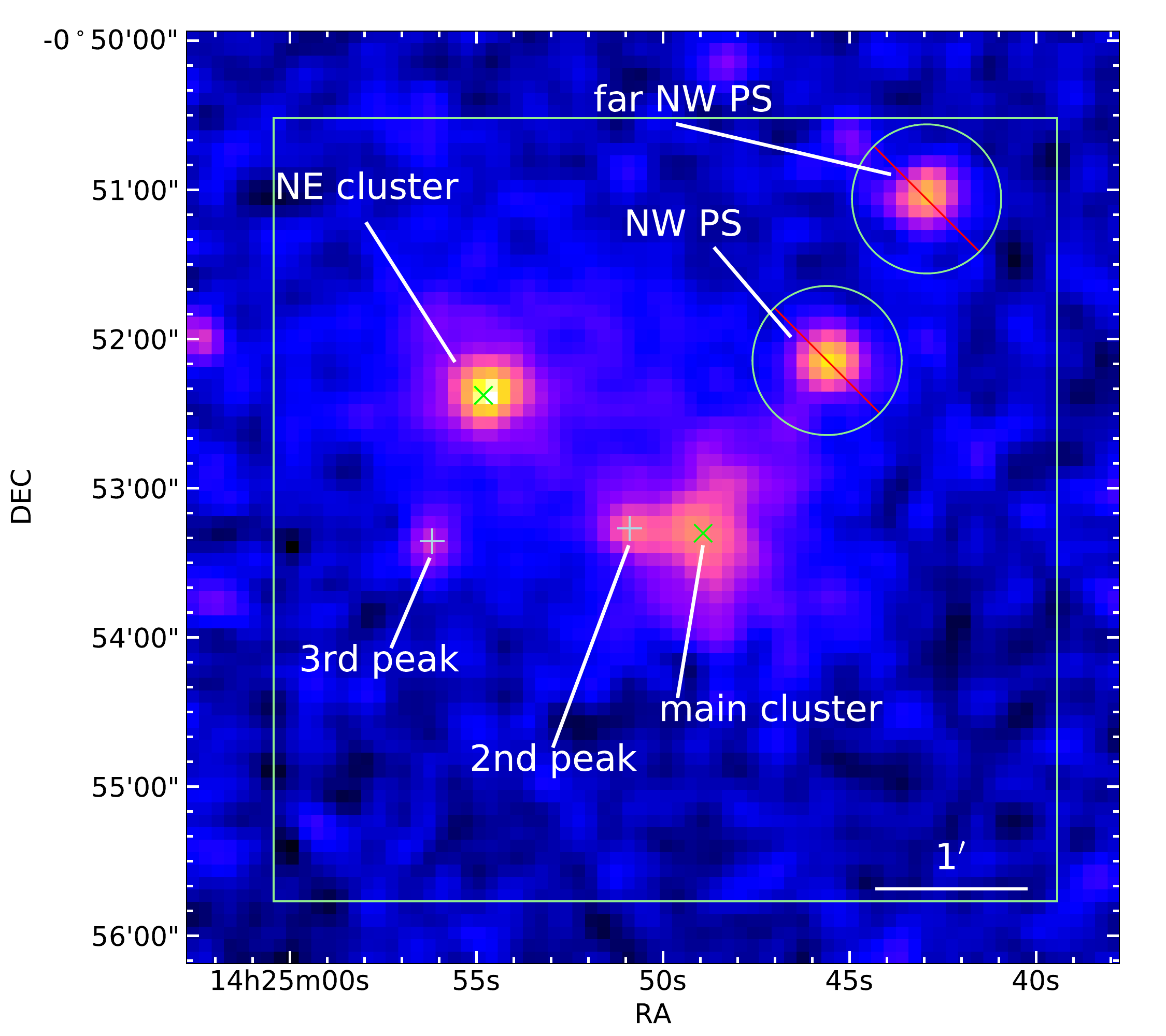}
\caption{X-ray surface brightness distribution near the {\it Eye of Horus}. It is a merged image of MOS1, 2 and pn in the 0.4--2.3 keV band, after subtracting the non-X-ray background (NXB), correcting exposure-time, and smoothing with $\sigma=5''$. The green lines show the region for image fitting. The green and the blue cross points show the best-fit center positions of each source obtained from image fitting.}
\label{fig:x-ray_img}
\end{figure}

The {\it Eye of Horus} was observed with {\it XMM-Newton} on 2018 January 6. Table~\ref{tab:reduction} shows detail information on the observation. The medium filter for MOS \citep{MOS} and the thin filter for pn \citep{pn} were selected, respectively. We used SAS version 17.0.0 to extract images and spectra, and CCF version XMM-CCF-REL-356 as a calibration database. We used the data analysis pipeline of \citet{XCASE} to adopt the standard data reduction. Raw data were filtered with the standard method and the events that matched the conditions of $\rm FLAG = 0$ and $\rm PATTERN \leq 12$ for MOS, and $\rm FLAG = 0$ and $\rm PATTERN \leq 4$ for pn were extracted. Bright point sources were excluded by using \verb+cheese+ command.\\
\begin{table*}
\begin{center}
\caption{Observation log. The flare time is removed from the exposure time.}
\label{tab:reduction}
\begin{tabular}{cccccc}
\hline\hline
Target name& Obs ID&Start time [UT] & Exposure time (MOS1 / MOS2 / pn) [ks] & RA & Dec \\\hline
HSC J142449-0053&0800790101&2018-01-06 06:26:13&43.4 / 44.5 / 30.0&14:24:49.0&-00:53:21.6\\
\hline
\end{tabular}
\end{center}
\end{table*}
\indent Fig.~\ref{fig:x-ray_img} shows a composite image of MOS1, 2 and pn in the 0.4--2.3 keV band. 
In addition to the main cluster and the NE cluster, a couple of sources were detected. There are two very bright point sources, $\sim 1'$ and $\sim 2'$ northwest of the main cluster (hereafter NW PS and far NW PS). There are two faint point sources, $\sim 30''$ east of the main cluster and $\sim 1'$ south-southeast of the NE cluster, respectively (hereafter the 2nd--peak and the 3rd--peak).

\section{Imaging analysis}
\label{sec:img}
We fitted the X-ray image to determine the center coordinates of the main cluster and the NE cluster. A fitting area of 315$''$ square, roughly centered at the main cluster, was selected to evaluate background properly as shown in Fig.~\ref{fig:x-ray_img}. The NW PS and the far NW PS are brighter than the main cluster, so the circular regions with a radius of 30$''$ centered at the point sources were excluded. 

\indent We used Sherpa \citep{Sherpa} in CIAO version 4.10 \citep{CIAO} for image fitting. 
The source distribution models were convolved with the PSF model and the exposure maps.
The PSF maps were generated at the energy of 1.35 keV and at the on-axis position using \verb+psfgen+ tool. The images, the PSF maps, and the exposure maps were rebinned so that the pixel size becomes $5''$. 
The three images were fitted simultaneously using the C statistic \citep{Cash}.

\indent We adopted a 2-dimensional single-$\beta$ model for the main cluster, a 2-dimensional double-$\beta$ model for the NE cluster, a 2-dimensional delta function for the 2nd--peak and the 3rd--peak, respectively, and a constant for the background component. We set the central coordinates of each source model as free parameters. The parameters of the corresponding models were linked with each other among the detectors, except for the normalizations. Two $\beta$'s of the double-$\beta$ model used for the NE cluster were also linked. The normalizations of the corresponding models were constrained so that the ratio becomes the same among the detectors.

\indent Table~\ref{tab:img_fit} shows the result of the image fitting. The center coordinates of each cluster were obtained with an accuracy of $\sim1''$ and each parameter was converged to an appropriate value as a cluster.
Note that, when we removed the constraints of the normalizations, $\beta$ of the NE cluster became relatively small ($\beta=0.46^{+0.10}_{-0.02}$) compared with typical clusters \citep[$\beta=0.64\pm0.32$,][]{Ota_Mituda2004}. Also note that $\beta$ became inappropriate when we unlinked the two $\beta$ of the double $\beta$-model or change the double $\beta$-model to the single $\beta$-model at the NE cluster.
\begin{table*}
\centering
\caption{Best--fit parameters obtained from the image fitting.}
\label{tab:img_fit}
\begin{tabular}{ccccccc}\\\hline
Parameter&Main cluster&NE cluster inner&NE cluster outer&2nd--peak&3rd--peak&Background\\\hline\hline
\rule[3pt]{0pt}{8pt}$r_c$ [arcsec / kpc]&$21.5^{+4.1}_{-1.9}~/~194^{+31}_{-15}$&$2.27^{+0.62}_{-0.16}~/~17.1^{+4.7}_{-1.2}$&$60.0^{+9.7}_{-5.6}~/~451^{+73}_{-42}$&-&-&-\\
\rule[3pt]{0pt}{8pt}$\beta$&$0.67^{+0.08}_{-0.03}$&\multicolumn{2}{c}{$0.78^{+0.14}_{-0.06}$}&-&-&-\\
\rule[3pt]{0pt}{8pt}$\rm{RA~[deg]}$&$216.20386^{+0.00030}_{-0.00029}$&\multicolumn{2}{c}{$216.22837^{+0.00012}_{-0.00011}$}&$216.21185^{+0.00108}_{-0.00031}$&$216.23378^{+0.00136}_{-0.00002}$&-\\
\rule[3pt]{0pt}{8pt}$\rm{DEC~[deg]}$&$-0.89034^{+0.00027}_{-0.00029}$&\multicolumn{2}{c}{$-0.87489^{+0.00014}_{-0.00016}$}&$-0.89053^{+0.00077}_{-0.00061}$&$-0.89204^{+0.00089}_{-0.00050}$&-\\
\rule[3pt]{0pt}{8pt}$\rm{norm_{~MOS1} [cnt/s/deg^2]}$&$11.6^{+0.7}_{-0.4}$&$518^{+466}_{-167}$&$2.0\pm0.3$&$82\pm16$&$59\pm10$&$0.60\pm0.03$\\
\rule[3pt]{0pt}{8pt}$\rm{norm_{~MOS2}~[cnt/s/deg^2]}$&$13.3^{+1.0}_{-0.9}$&\multicolumn{5}{c}{$\rm =norm(MOS2,main~cluster)\times norm(MOS1,each~component)~/~norm(MOS1,main~cluster)$}\\
\rule[3pt]{0pt}{8pt}$\rm{norm_{~pn}~[cnt/s/deg^2]}$&$12.7^{+0.8}_{-0.5}$&\multicolumn{5}{c}{$\rm =norm(pn,main~cluster)\times norm(MOS1,each~component)~/~norm(MOS1,main~cluster)$}\\
\hline
\multicolumn{1}{c}{C-statistic}&\multicolumn{6}{c}{16011}\\
\multicolumn{1}{c}{d.o.f.}&\multicolumn{6}{c}{11208}\\
\multicolumn{1}{c}{C-statistic/d.o.f.}&\multicolumn{6}{c}{1.429}\\\hline
\end{tabular}
\end{table*}

\begin{figure}
\includegraphics[width=\columnwidth]{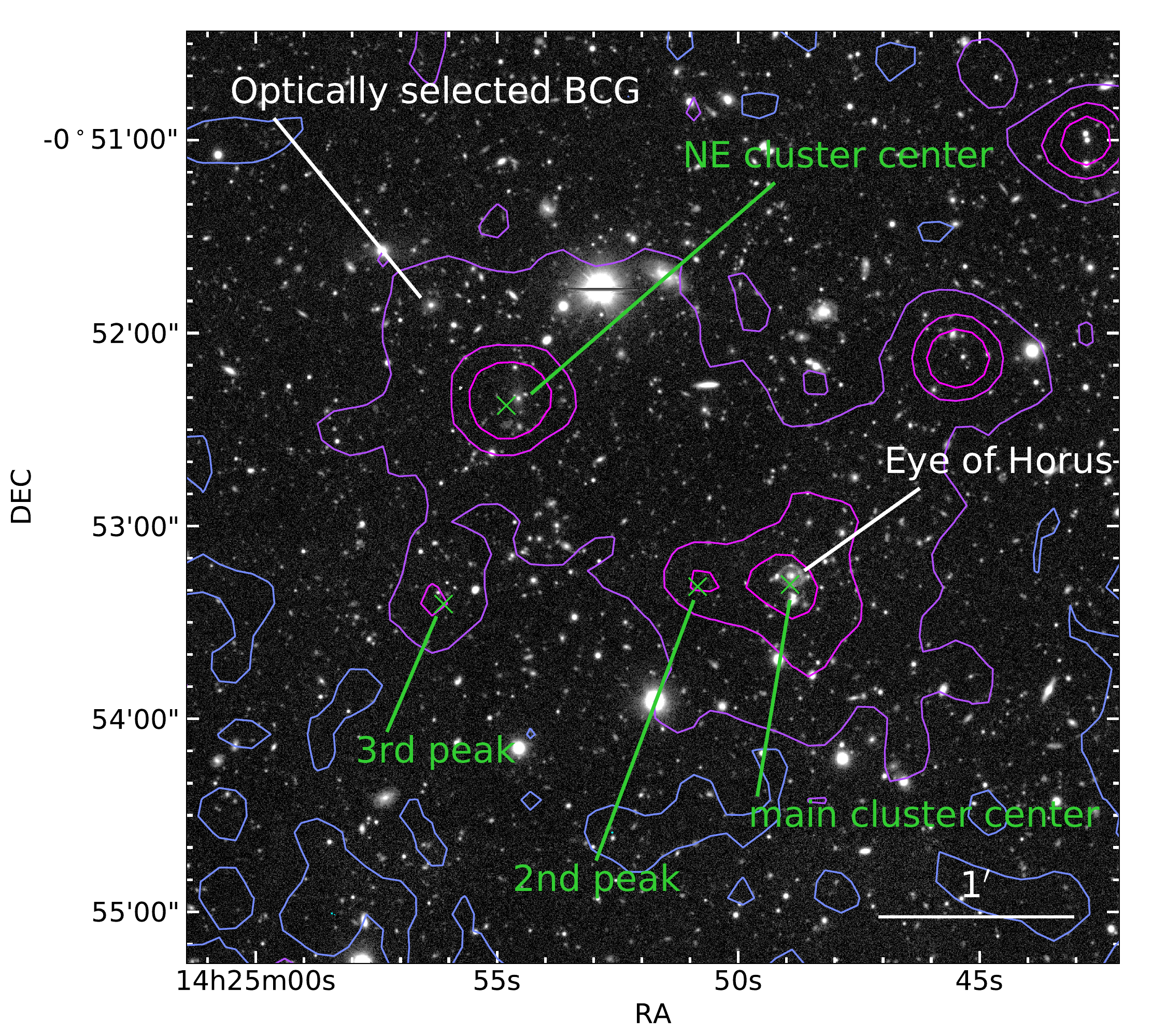}
\caption{X-ray contours generated from Fig.~\ref{fig:x-ray_img}, superposed on the optical image observed with Subaru--HSC. Green crosses show the best-fit center positions of the main cluster, the NE cluster, the 2nd peak, and the 3rd peak.}
\label{fig:HSC-XMM}
\end{figure}

\indent Fig.~\ref{fig:HSC-XMM} shows an X-ray contour map superposed on the optical image. The positions of the X-ray sources are denoted by the green crosses. The X-ray peak of the main cluster is consistent with the optical position of the {\it Eye of Horus}. It is also consistent with the result of CAMIRA cluster finding algorithm that HSC J142449-005322 is the central galaxy of the main cluster. But the center position determined from image fit is $3.8\pm1.0''$ shifted to the south from the center of the {\it Eye of Horus}. On the other hand, the best-fit position of the NE cluster deviated considerably ($\sim 40''$) from the optically selected BCG (see Fig.~\ref{fig:HSC-XMM}) ,which is probably due to an error of optical identification, since CAMIRA cluster finding algorithm has a certain uncertainty in the determination of the BCG. There is an elliptical galaxy SDSS J142454.68-005226.3, $3.8\pm 0.5''$ northwest of the X-ray peak of the NE cluster, and we infer that this object is the correct BCG of the NE cluster. The X-ray peak of the main cluster and the NE cluster shifted 3.8$\pm1.0''$ south and 3.8$\pm0.5''$ southeast from the optical counterpart, respectively. These discrepancies are higher than the position uncertainty of the EPIC, which is 1.2$''$ (1$\sigma$) according to XMM-Newton Calibration Technical Notes v3.11\footnote{http://xmm2.esac.esa.int/docs/documents/CAL-TN-0018.pdf}, and we infer that these discrepancies are caused by the cluster-scale structure.

\begin{figure}
\includegraphics[width=\columnwidth]{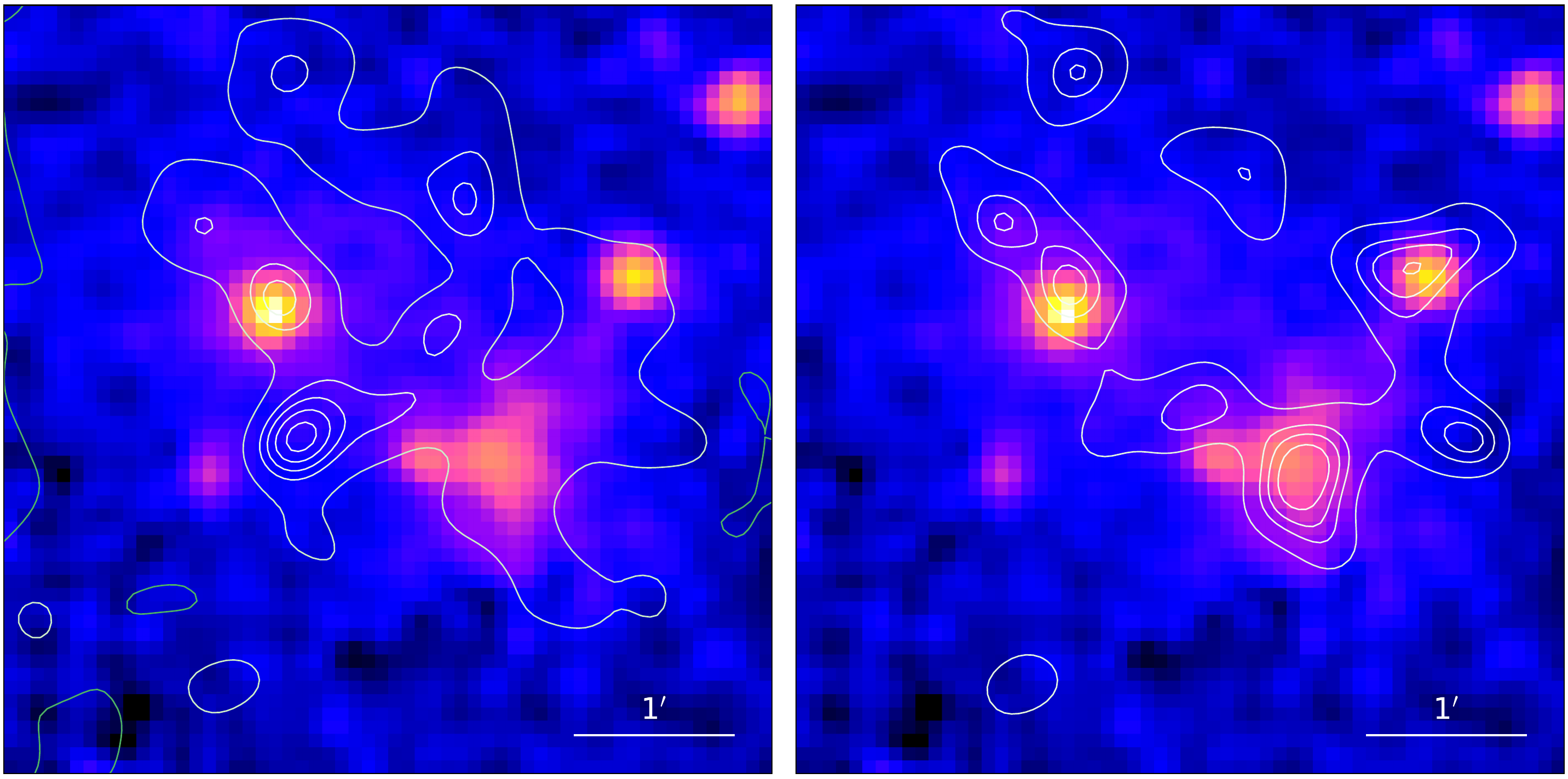}
\caption{Galaxy probability density distribution contours at $0.762<z<0.774$ (left) and $0.795<z<0.807$ (right), overlaid with the X-ray image. The redshifts correspond to the NE cluster and the main cluster, respectively. The X-ray image is the same as Fig.~\ref{fig:x-ray_img}}
\label{fig:gal_map-xray}
\end{figure}

\begin{figure}
\includegraphics[width=\columnwidth]{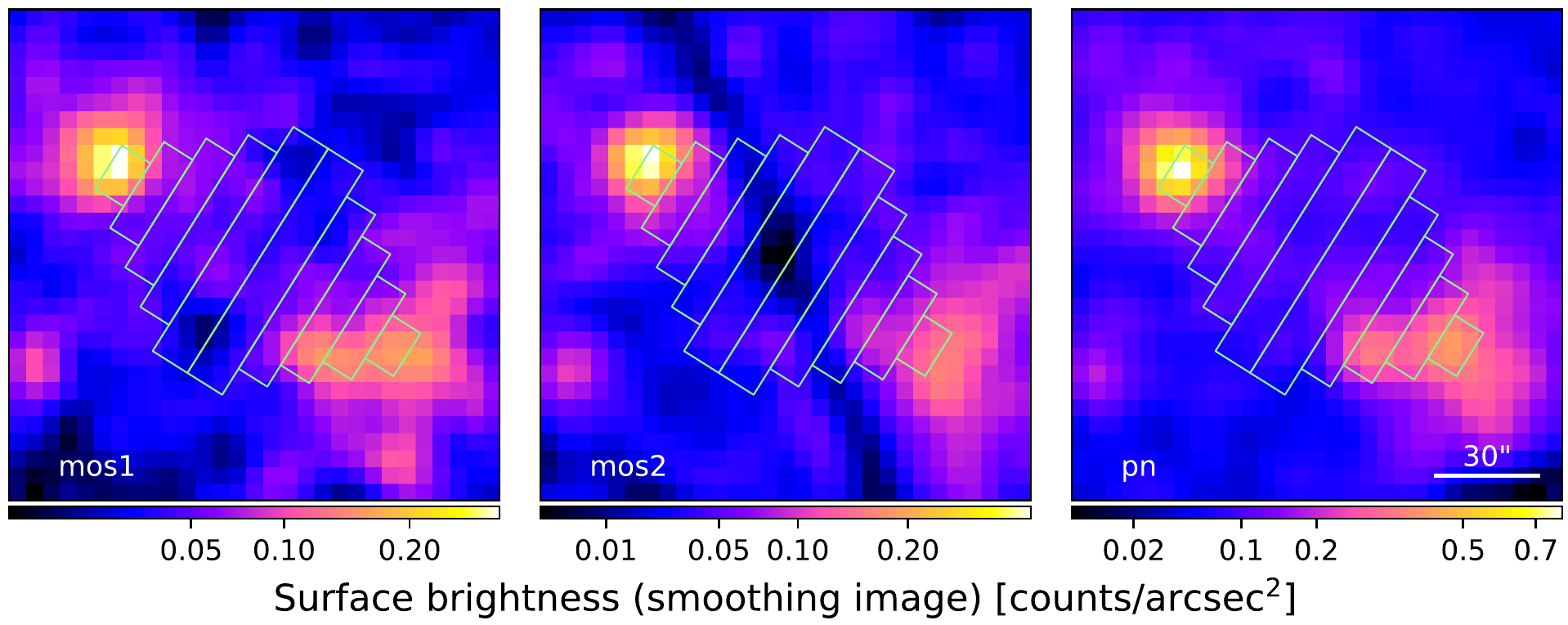}
\\
\includegraphics[width=\columnwidth]{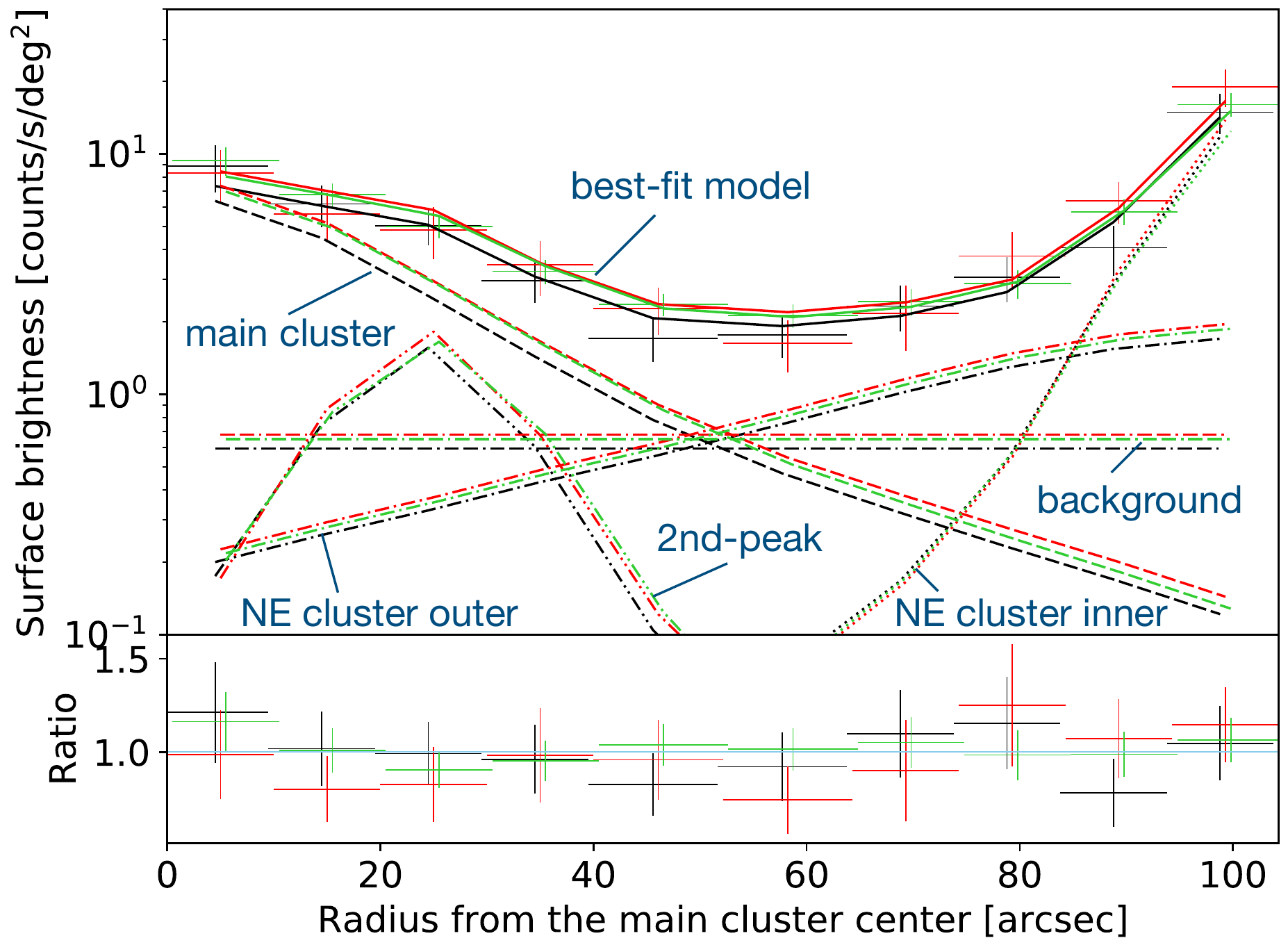}
\caption{The X-ray surface brightness profile between the main cluster and the NE cluster (brack: MOS1, red: MOS2, green: pn). The top panels show extracted regions of X-ray surface brightness. The dashed, dotted, dash-dotted, dot-dot-dashed, dash-dash-dotted lines correspond to the main cluster, the NE cluster outer, the NE cluster inner, the 2nd--peak, the background components of the best-fit model, respectively, and the solid line shows the sum of them.}
\label{fig:SB_profile}
\end{figure}

\indent We investigated whether the main cluster and the NE cluster are a merger. 
Fig.~\ref{fig:gal_map-xray} shows a contour map of the galaxy probability density distribution in the redshift band of each cluster generated by MIZUKI \citep{MIZUKI}, overlaid with the X-ray image. MIZUKI determines the probability density distribution of the galaxies at any redshift from the photometric data. From the left panel, in the redshift band of the NE cluster, the galaxy density is relatively low around the main cluster, and there are some peaks of the galaxy density around the NE cluster. However, from the right panel, these peaks around the NE cluster also exist in the redshift band of the main cluster. Therefore, it is not clear from photometric data whether these two clusters are a merger or not.
Fig.~\ref{fig:SB_profile} shows the X-ray surface brightness profile between the main cluster and the NE cluster. When the two clusters are a merger, it is known that excess of X-ray emission is observed between two clusters, due to compression of ICM by the merging \citep[e.g., between A399 and A401,][]{A399A401_Fujita96}. However, there is no significant excess emission beyond the error range over the sum of the two cluster models between the main cluster and the NE cluster. This result indicates that these two clusters are likely not a merger.

\section{Spectral analysis}
\label{sec:spec}
\begin{figure}
\includegraphics[width=\columnwidth]{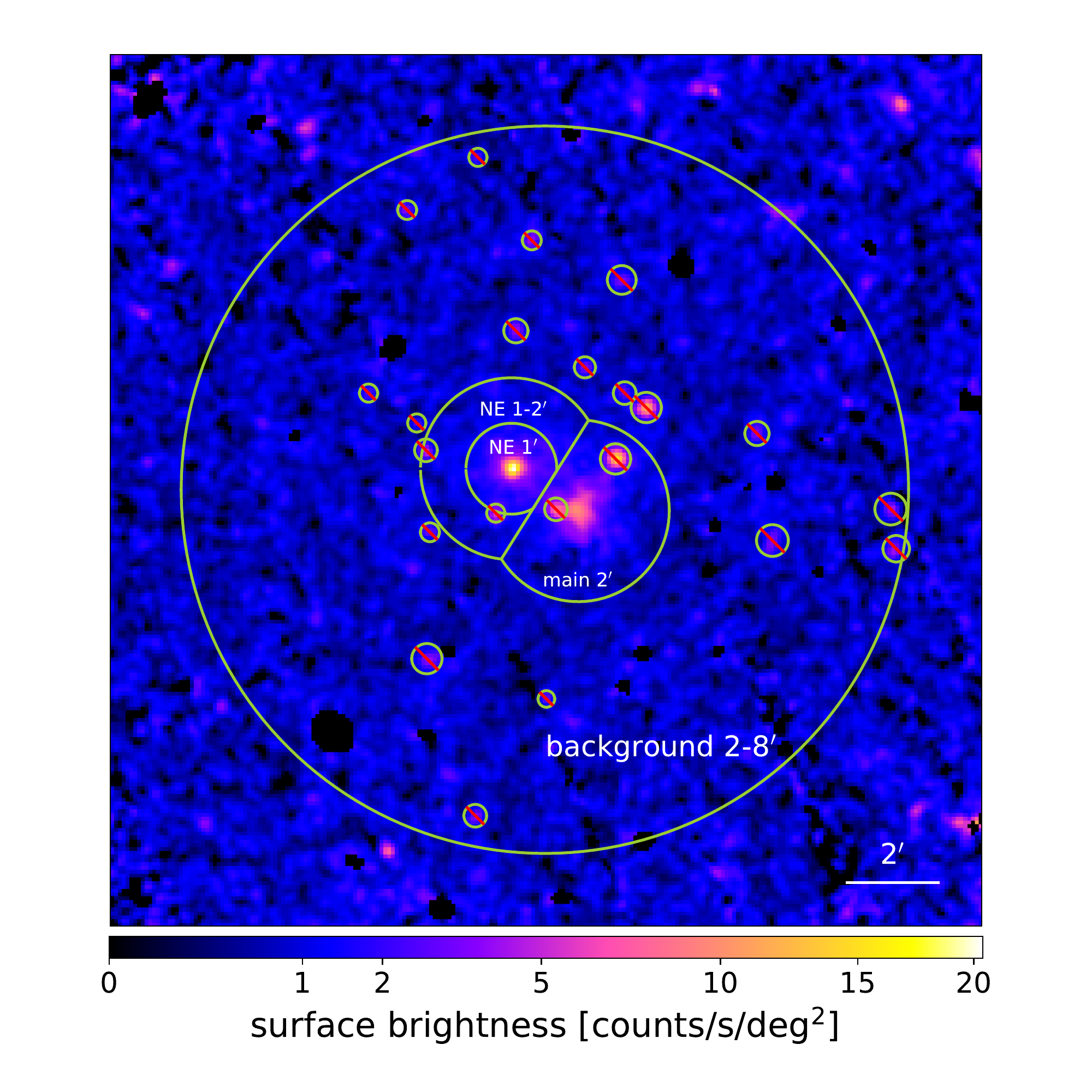}
\caption{Source and background regions for spectral fitting. The image is the same as Fig.~\ref{fig:x-ray_img}}
\label{fig:fit_reg}
\end{figure}

We fitted the EPIC spectra to determine the ICM temperatures of the two clusters. 
Fig.~\ref{fig:fit_reg} shows the definition of the spectral integration regions. From the imaging analysis, we found that the intensity of cluster emission reaches the background level at the radius of $\sim$100$''$ (764 and 752 kpc for the main and the NE cluster, respectively). Therefore, with a margin, let the radius 2$'$ be the boundary between each cluster region and the background region. Since the main cluster and the NE cluster are separated only by $\sim$100$''$, we divided the regions of the main cluster and the NE cluster at the midpoint between them. The background region was defined as a circle of a radius of 8$'$ centered at the midpoint, excluding the main cluster and the NE cluster regions. Note that, the NE cluster region was divided into the inner region ($r\leq1'$) and the outer region ($1'<r\leq2'$) to consider temperature gradient because the X-ray emission of the NE cluster is concentrated in the central region.
There are many point sources in these regions. Bright point sources were removed by the \verb+cheese+ command. Faint point sources that can be distinguished by eye, but cannot be removed by \verb+cheese+ command were manually removed. Table~\ref{tab:reg_info} summarizes photon counts of the four regions.
\begin{table}
\centering
\caption{Photon counts in the four regions. The energy bands are 0.3--11 keV for the MOS spectra and 0.4--11 keV for the pn spectra, respectively. The net count means the QPB-subtracted count.}
\label{tab:reg_info}

\begin{tabular}{ccccc}\\\hline
Region&Instrument&Counts&Net counts\\\hline\hline
Main cluster&MOS1&1683&1129.1\\
&MOS2&1678&1115.3\\
&pn&3560&2022.9\\\hline
NE cluster inner&MOS1&685&571.3\\
&MOS2&725&577.7\\
&pn&1692&1297.3\\\hline
NE cluster outer&MOS1&932&490.5\\
&MOS2&804&494.1\\
&pn&1183&815.4\\\hline
Background&MOS1&17059&7703.5\\
&MOS2&21771&8985.0\\
&pn&41210&19698.7\\\hline
\end{tabular}
\end{table}

\indent Quiescent particle background (QPB) spectra were generated using \verb+mos+\_\verb+back+ and \verb+pn+\_\verb+back+ and subtracted from the spectra of the four regions. Then the spectra were rebinned so that each bin contained at least 20 counts. The energy band was set to 0.3--11.0 keV for MOS and 0.4--11.0 keV for pn.\\
\indent We used XSPEC version 12.10.1 \citep{XSPEC} for spectral fitting. We adopted the APEC version 3.0.9 \citep{APEC} as an optically thin thermal plasma model. We adopted \verb+phabs+ as a photoelectric absorption model, and the column density was fixed at 3.25$\times10^{20}$ cm$^{-2}$ based on the LAB survey \citep[]{LABsurvey}. EPIC spectra have many background components \citep{MOS_background1,MOS_background2,pn_background,XMM_SWCX1,XMM_SWCX2}. 
We fitted all the spectra of the detectors and the four regions simultaneously following \citet{Snowden}. 

\indent Emission of each cluster was represented by \verb+phabs*apec+, and the parameters were linked among detectors. The redshift of each cluster was fixed at the value derived from photometric data \citep{CAMIRA_list}. The abundances of the NE cluster component in the inner and the outer regions were linked with each other. The projection effects are considered, assuming that each source is spherically symmetric with constant density.

\indent The response files were created by SAS tools. RMFs were generated together with spectra using \verb+mos-spectra+ and \verb+pn-spectra+ tools. 
We adopted extended source ARFs that were generated using \verb+arfgen+ tool. We adopted $\chi^2$ statistic to the statistical test.

\begin{figure}
\includegraphics[width=\columnwidth]{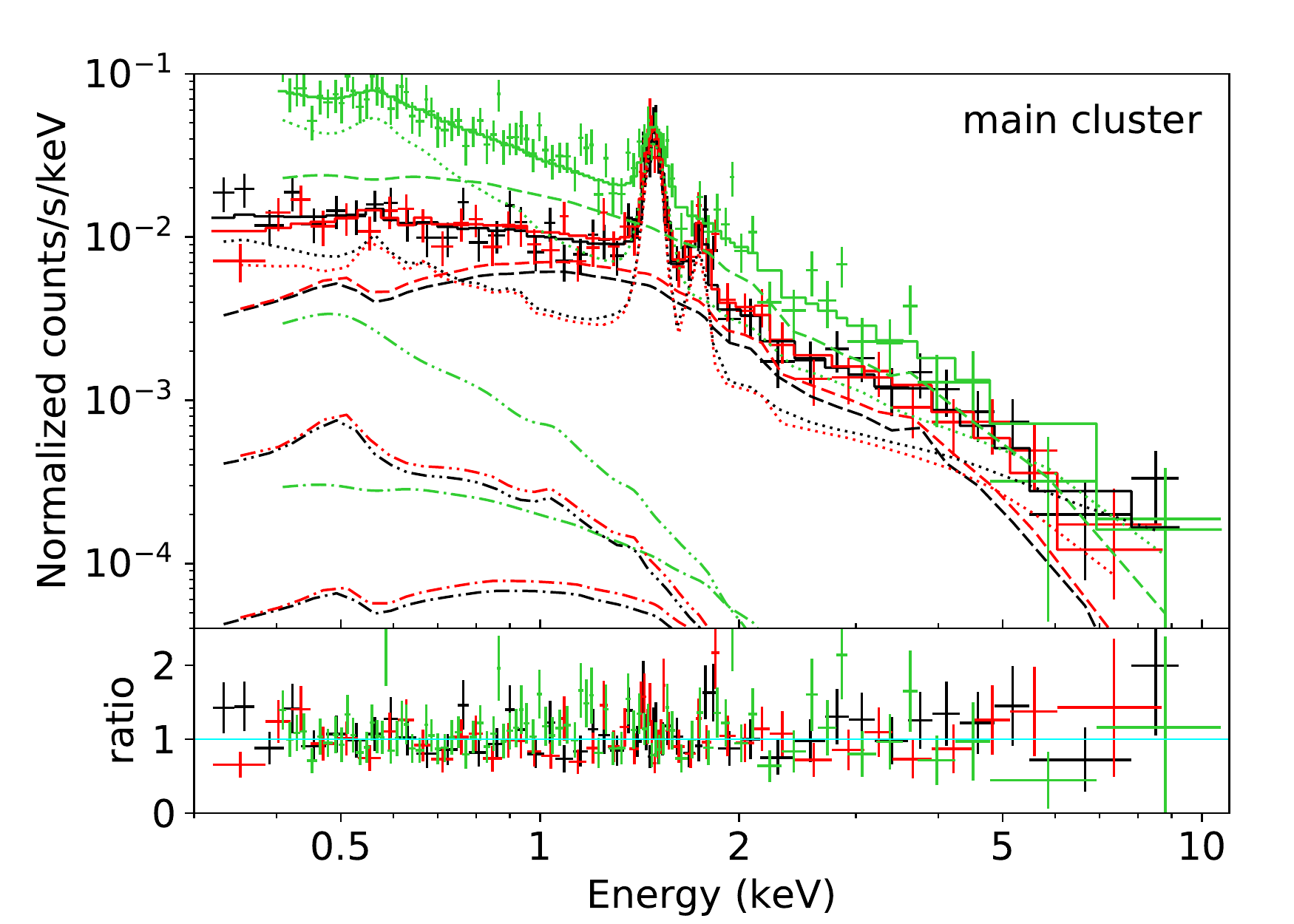}
\includegraphics[width=\columnwidth]{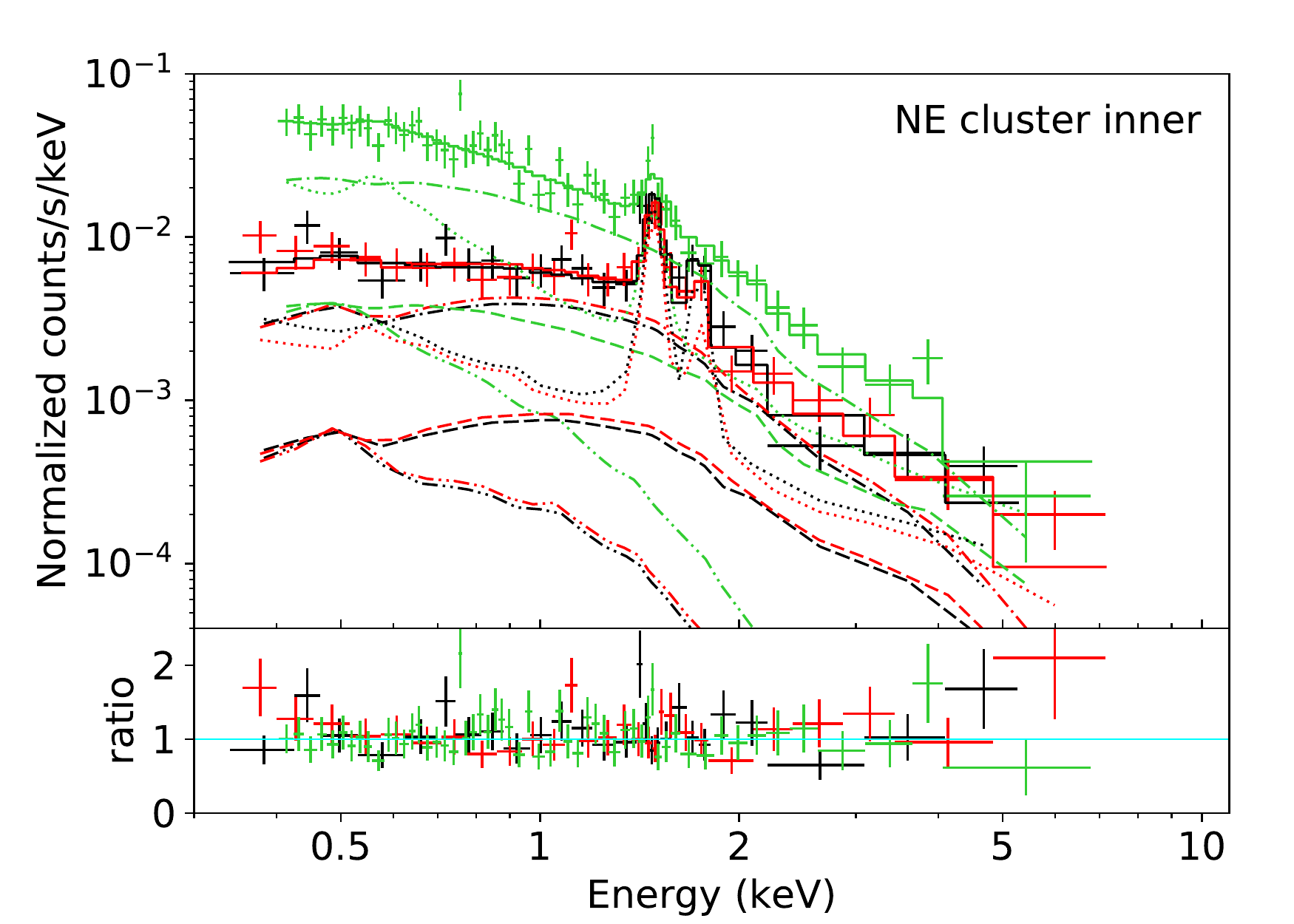}
\includegraphics[width=\columnwidth]{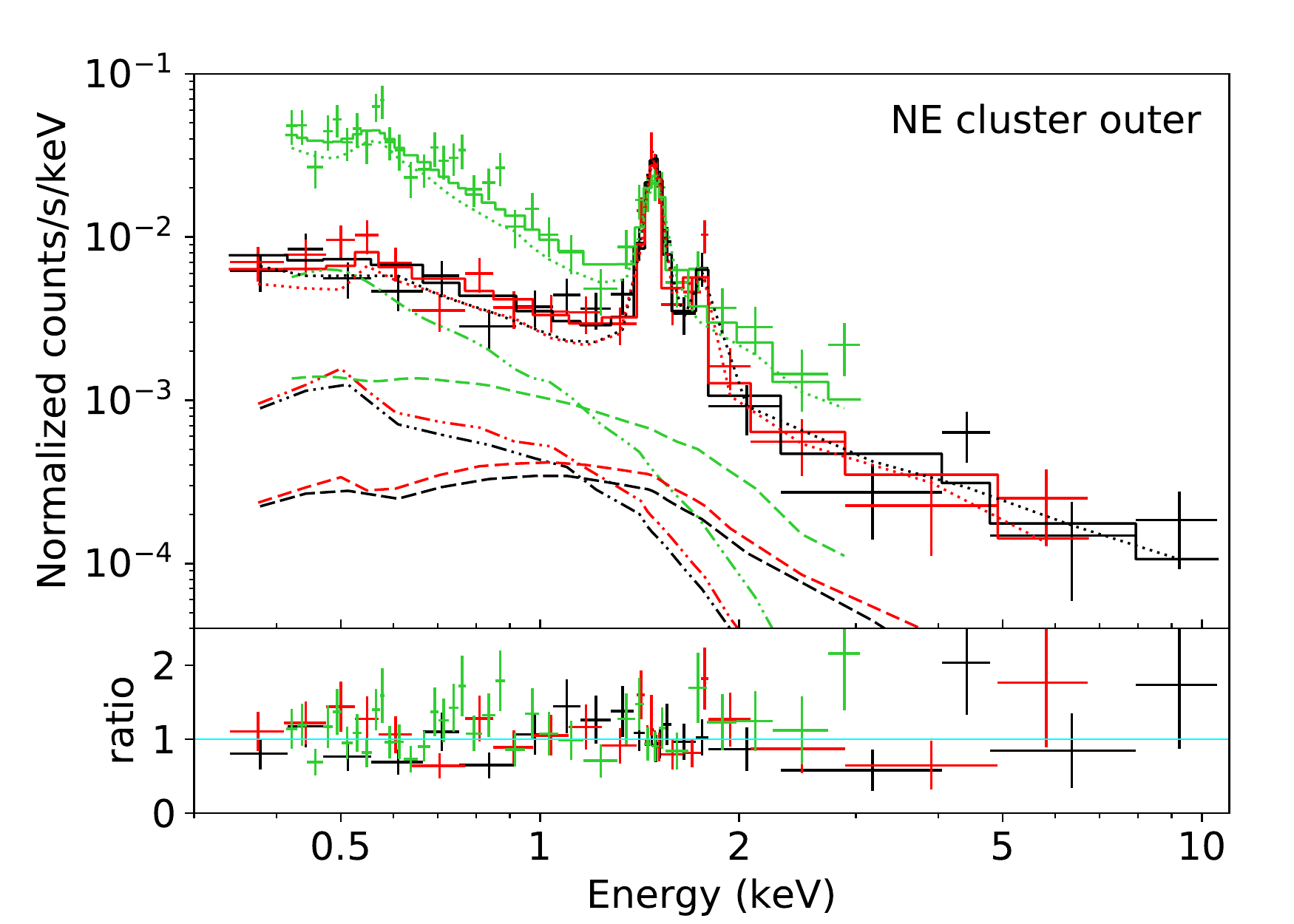}
\caption{QPB-subtracted spectra and best-fit models of the three source regions (black: MOS1, red: MOS2, green: pn). The dashed, dot-dashed, dot-dot-dashed and dotted lines show the main cluster, the NE cluster inner, the NE cluster outer and the background component, respectively.}
\label{fig:spec}
\end{figure}

\indent Fig.~\ref{fig:spec} and Table~\ref{tab:spec_fit} show the best-fit parameters of the spectral fitting.
From Table~\ref{tab:spec_fit}, the temperature of the NE cluster has radial dependence although it is marginal.
We also tested whether there is a temperature gradient in the main cluster, by dividing the region of the main cluster into two, an inner circle of 1$'$ radius and an annulus of 2$'$ outer radius, but we could not find the difference in the temperature between the two regions ($kT_{\rm in}=5.6^{+0.8}_{-0.7}$~keV, $kT_{\rm out}=5.9^{+14}_{-2.8}$~keV). 
When the redshifts were made free, they became unacceptably high ($z_{\rm main}=2.71^{+0.13}_{-0.08}$, $z_{\rm NE}=3.71^{+0.24}_{-0.12}$).
Note that, since the instrumental Al K$\alpha$ and Si K$\alpha$ lines are predominant in the 1.3--1.9 keV range in MOS spectra and in the 1.3-1.6 keV range in pn spectra (see Fig~\ref{fig:spec}), we evaluated the influence of these lines to the fitting result by excluding these energy ranges. The parameters of the ICM unchanged within the error ranges of Table \ref{tab:spec_fit}.
\begin{table}
\centering
\caption{Results of the spectral fitting of the cluster regions.}
\label{tab:spec_fit}
\begin{tabular}{cccc}\\\hline
parameter&Main cluster&NE cluster inner&NE cluster outer\\\hline\hline
\rule[3pt]{0pt}{8pt}kT [keV]&\multicolumn{1}{c}{$6.07^{+1.22}_{-0.89}$}&\multicolumn{1}{c}{$3.30^{+0.47}_{-0.40}$}&\multicolumn{1}{c}{$0.94^{+0.87}_{-0.32}$}\\
\rule[3pt]{0pt}{8pt}Abundance [$Z_\odot$]&\multicolumn{1}{c}{$0.21^{+0.35}_{-0.21}$}&\multicolumn{2}{c}{$0.06^{+0.25}_{-0.06}$}\\
\rule[3pt]{0pt}{8pt}Redshift&\multicolumn{1}{c}{0.80094 (fixed)}&\multicolumn{2}{c}{0.76828 (fixed)}\\
\hline
\multicolumn{1}{c}{d.o.f.}&\multicolumn{3}{c}{1284}\\
\multicolumn{1}{c}{$\chi^2$}&\multicolumn{3}{c}{1625.04}\\
\multicolumn{1}{c}{Reduced-$\chi^2$}&\multicolumn{3}{c}{1.266}\\\hline
\end{tabular}
\end{table}

From the result of image fitting and spectral fitting, the surface brightness distribution of the NE cluster is well represented by a double-$\beta$ model, and the spectrum is represented by a two -temperature model. These results suggest that the NE cluster is divided into two components, the inner component ($r_c=2.44^{+0.60}_{-0.16}$ arcsec, $kT=3.30^{+0.47}_{-0.40}$~keV) and outer component ($r_c=57.8^{+9.1}_{-5.5}$ arcsec, $kT=0.94^{+0.87}_{-0.32}$~keV), and the gravitational potential of the NE cluster is concentrated in the central elliptical galaxy.

\section{Discussion}
\subsection{Mass of the clusters}
Based on the distributions and the temperatures of the ICM obtained by the results of the image fitting and the spectral fitting, we calculated the hydrostatic mass of the main cluster and the NE cluster, and evaluated the effect of the cluster-scale environment on the lens image of the {\it Eye of Horus}. When the ICM is in hydrostatic equilibrium and its surface brightness distribution follows a $\beta$-model, the hydrostatic mass density at a radius $r$ from the cluster center can be represented as follows \citep[]{Ota1998}
\begin{eqnarray}
\rho (r)=\frac{1}{4\pi r^2}\cdot\frac{3k_BT\beta}{G\overline{m}}\left\{\frac{3r^2}{r^2+r_c^2}-\frac{2r^4}{\left(r^2+r_c^2\right)^2}\right\},
\label{eq:rho}
\end{eqnarray}
where $k_B$, $\overline{m}$, and $G$ are the Boltzmann constant, the average ion mass, and the gravitational constant, respectively. 
From this equation, the hydrostatic mass which are projected on the sky plane within a radius $r$ is given by
\begin{eqnarray}
 M(r)=\frac{3k_{B}T\beta r}{G\overline{m}}\cdot\frac{r^2}{r^2+r_c^2}.
\label{eq:mhe}
\end{eqnarray}
$r_{200}$ ($r_{500}$) is defined as the radius within which the average density is 200 (500) times the critical density, and  $M_{200}$ ($M_{500}$) is the hydrostatic mass within $r_{200}$ ($r_{500}$), i.e., 
\begin{eqnarray}
M_{200}\equiv M(r_{200})=200\cdot\frac{3H^2}{8\pi G}\cdot\frac{4\pi}{3}\cdot r_{200}^3,
\label{eq:m200}
\end{eqnarray}
and
\begin{eqnarray}
M_{500}\equiv M(r_{500})=500\cdot\frac{3H^2}{8\pi G}\cdot\frac{4\pi}{3}\cdot r_{500}^3,
\label{eq:m500}
\end{eqnarray}
where $H$ is the Hubble parameter, $H=106~{\rm km~s^{-1}~Mpc^{-1}}$ for the main cluster and $H=104~{\rm km~s^{-1}~Mpc^{-1}}$ for the NE cluster, respectively. Using equations (\ref{eq:mhe}) and (\ref{eq:m200}) or (\ref{eq:m500}), $r_{200}$ and $M_{200}$, or $r_{500}$ and $M_{500}$, can be determined. 
Since the NE cluster has two components, when we use equation (\ref{eq:mhe}) for the NE cluster, we combined the inner component and the outer component with their density ratio.
 
\indent Table \ref{tab:M200M500_xray} shows $M_{200}$, $M_{500}$ and $r_{200}$, $r_{500}$ thus obtained. We confirmed that there are two massive cluster, $5.6^{+1.3}_{-0.8}\times10^{14}~M_{\odot}$ and $2.2^{+0.5}_{-0.3}\times10^{14}~M_{\odot}$, respectively, in the field of the {\it Eye of Horus}. The cluster mass inferred from the mass-richness scaling relation \citep{Ncor_mass} is also shown in Table \ref{tab:M200M500_richness} for comparison.
The two clusters have comparable optical mass, while the hydrostatic mass of the main cluster estimated by the X-ray data is almost three times larger than that obtained from the mass-richness relation. This is probably because measurement of the number of the member galaxies has large uncertainty in the distant cluster, and we assumed spherical symmetry and extrapolated the isothermal single-$\beta$ model to the outside where X-ray emission cannot be seen. 

\begin{table}
\centering
\caption{$M_{200}$ and $M_{500}$ of the main cluster and the NE cluster determined by the X-ray data.}
\label{tab:M200M500_xray}
\begin{tabular}{ccc}\\\hline
source&Main cluster&NE cluster\\\hline\hline
\rule[3pt]{0pt}{8pt}$r_{200}$ (arcsec / Mpc)&168 / 1.28&127 / 0.95\\
\rule[3pt]{0pt}{8pt}$r_{500}$ (arcsec / Mpc)&105 / 0.80&80 / 0.60\\
\rule[3pt]{0pt}{8pt}$M_{200}$ ($M_\odot$)&$5.6^{+1.3}_{-0.8}\times10^{14}$&$2.2^{+0.5}_{-0.3}\times10^{14}$\\
\rule[3pt]{0pt}{8pt}$M_{500}$ ($M_\odot$)&$3.4^{+0.8}_{-0.5}\times10^{14}$&$1.4^{+0.3}_{-0.2}\times10^{14}$\\\hline
\end{tabular}
\end{table}

\begin{table}
\centering
\caption{$M_{200}$ and $M_{500}$ of the main cluster and the NE cluster inferred from the mass-richness scaling.}
\label{tab:M200M500_richness}
\begin{tabular}{ccc}\\\hline
source&Main cluster&NE cluster\\\hline\hline
$N_{\rm gal}$&34&37\\
\rule[3pt]{0pt}{8pt}$M_{200}$ ($M_\odot$)&$2.2\times10^{14}$&$2.5\times10^{14}$\\
\rule[3pt]{0pt}{8pt}$M_{500}$ ($M_\odot$)&$1.4\times10^{14}$&$1.6\times10^{14}$\\\hline
\end{tabular}
\end{table}

\subsection{Mass within the Einstein radius}
The total mass $M_{\rm tot}$ projected on the sky plane within the Einstein radius can be calculated from the Einstein radius of the lens galaxy, and angular diameter distances to the lens galaxy, to the background source, and between the lens galaxy and the background source. The {\it Eye of Horus} has two background galaxies \citep[S1 and S2 in][]{MTanaka} and Einstein radii. The Einstein radius of S2 is affected not only by the lens galaxy but also by S1 since S2 is located behind S1. Therefore, in this paper, we only used S1. Since the Einstein radius of the S1 is $2.14\pm0.02$ arcsec \citep{Sonnenfeld2019}, $M_{\rm tot}$ is $(2.98\pm0.04)\times10^{12}~M_\odot$, which is $\sim$4.5 times the stellar mass of the lens galaxy \citep[$6.6^{+0.7}_{-0.1}\times10^{11}M_\odot$,][]{MTanaka}. 

\indent Using equation (\ref{eq:rho}), we calculated $M_{\rm E}$, the mass of each cluster projected on the sky plane within the Einstein radius of the lens galaxy, which is summarized in Table \ref{tab:M_E}.
In calculating $M_{\rm E}$ for each cluster, we extrapolated the result of the image and spectral fitting to a sufficiently distant point along the line of sight where the mass calculation converges, although the X-ray emission can be seen only up to $\sim r_{500}$.
Note that, we considered a statistical error ($1.0''$) and XMM position uncertainty ($1.2''$) as a position error.

\indent $M_{\rm E}$ of the main cluster is 
$1.1^{+0.7}_{-0.4}\times10^{12} M_{\odot}$, which is larger than the stellar mass of the lens galaxy, yet explains only
$25-60\%$ of $M_{\rm tot}$. Therefore, we investigated the cause of the discrepancy. First, we tested an NFW density profile \citep{NFWprofile} for the mass estimation because it has steeper and deeper potential at the central region than that of a single-$\beta$ model. 
We used relation between the scale radius of the NFW density profile $r_s$ and the core radius of the $\beta$ model $r_c$, $r_s=r_c/0.22$ \citep{Makino1998}. Then, we calculated $M_{\rm E}$ in the same manner and obtained 
$M_{\rm E,~NFW}=1.1^{+1.2}_{-0.5}\times10^{12}~M_{\odot}$. The upper limit is higher than the mass derived from the $\beta$ model. However, the mass discrepancy cannot be explained even if only the NFW density profile is adopted.
Therefore, 
we conclude that $M_{\rm E,~NFW}=1.1^{+1.2}_{-0.5}\times10^{12}~M_{\odot}$ is a secure limit of the mass related to the main cluster, 
and it can explain $21-76\%$ of $M_{\rm tot}$ as long as we estimate one-component gravitational potential. Note that, there is a possibility that the gravitational potential of the central lens galaxy was not measured properly since we used a single-$\beta$ model.
To evaluate the influence of the central lens galaxy, we tested to fit the X-ray surface brightness of the main cluster with a double-$\beta$ model. Then we calculated $M_E$ by stacking the components of the inner and the outer in the density ratio. However, we could not determine the mass since the normalization of the inner component has a large error ($M_{\rm E,~2\beta}=3.3^{+14.4}_{-2.4}\times10^{12}~M_{\odot}$).
We did not consider the radial dependence of the temperature for the main cluster, but the Einstein ring is much smaller than the core radius of the main cluster (25.0$''$), and is located completely within it. And the ICM of the main cluster is well represented by a one-temperature model. Therefore, the influence of the radial dependence of the temperature is expected to be small. 

\begin{table}
\centering
\caption{The result of the calculation of $M_{\rm E}$. $d_{\rm EoH}$ means projected distance between each cluster center position determined by X-ray image fitting and the lens galaxy of the {\it Eye of Horus}}
\label{tab:M_E}
\begin{tabular}{ccc}\\\hline
source&Main cluster&NE cluster\\\hline\hline
\rule[3pt]{0pt}{8pt}$d_{\rm EoH}$ (arcsec)&3.8&102\\
\rule[3pt]{0pt}{8pt}$M_{\rm E}$ ($M_\odot$)&$1.1^{+0.7}_{-0.4}\times10^{12}$&$6.1^{+1.4}_{-0.9}\times10^{10}$\\\hline
\end{tabular}
\end{table}

\indent Several studies reported that the hydrostatic mass is underestimated compared to $M_{\rm tot}$ derived from the strong lens data. \citet{Ota2004_CL0024+17} reported that the hydrostatic mass of the strong lens cluster CL 0024+17 is smaller by a factor of 2--3. They discussed possibilities that there are additional mass components since the lens cluster is a line of sight merger, or that there are substructures in the central region of the lens cluster, and the mass profile follows the NFW density profile rather than the $\beta$ model. 
Fig. \ref{fig:gal_map-xray} indicates that the main cluster has complex galaxy distribution along the line of sight of the {\it Eye of Horus}. There might be complex mass structures which are not considered in our analysis.
To evaluate the degree of anisotropy of the mass distribution, we divided the area into the sectors of an opening angle of $90^{\circ}$ in the NE, NW, SW and SE quadrants, and fitted the surface brightness distribution of each sector with a single-$\beta$ model. However, since there are not enough photon counts, we could not find any significant difference in each parameter. 
\citet{Hashimotodani1999} systematically studied 50 strong lens clusters and classified them into two types, one has the X-ray peak and the strong lens galaxy at the same center position, and the other at the different positions. He reported that the former and the latter follow $M_{\rm tot}/M_{\rm E}=2.17\pm 0.13$ and $M_{\rm tot}/M_{\rm E}=3.33\pm 0.39$, respectively. Since the X-ray peak of the main cluster is slightly shifted from the center of the lens galaxy, the {\it Eye of Horus} may follow the latter. 

\indent On the other hand, $M_{\rm E}$ of the NE cluster is only $\sim2\%$ of the $M_{\rm tot}$. 
This is probably because the gravitational potential of the NE cluster is concentrated in the central elliptical galaxy. 
Note that, however, several studies reported that the mass of the cluster whose center position is located far from the lens galaxy could affect the shape of lens image as external shear \citep[e.g., ][]{Grillo2008}. This effect needs to be considered in the detailed lens modeling of the {\it Eye of Horus} in future studies.

\section{Conclusions}
\label{sec:conc}
We observed X-ray emission around the {\it Eye of Horus} with {\it XMM--Newton} to evaluate the influence of cluster-scale mass structure on the lens image of the {\it Eye of Horus}. There are two clusters, the main cluster and the NE cluster which is located $\sim100''$northeast of the lens galaxy, and we found that the center position of the main cluster is located $3.8\pm1.0''$ south of the lens galaxy. The surface brightness distribution of the main cluster and the NE cluster is represented by a single, and a double $\beta$ model, respectively. We also revealed that the spectrum of the main cluster is represented by a one-temperature model, while that of the NE cluster needs a two-temperature model.

\indent The total mass projected on the sky plane within the Einstein radius $M_{\rm tot}$ determined by the Einstein radius is $\sim3.0\times10^{12}M_{\odot}$, which is $\sim 4.5$ times larger than the stellar mass of the lens galaxy. We calculated the hydrostatic mass projected on the sky plane within the Einstein radius of the lens galaxy, $M_{\rm E}$, using the X-ray data. $M_{\rm E}$ of the NE cluster is $6.1^{+1.4}_{-0.9}\times10^{10}~M_{\odot}$, which is only $\sim2\%$ of $M_{\rm tot}$. Therefore, the influence on $M_{\rm tot}$ is small. On the other hand, $M_{\rm E}$ of the main cluster is 
$1.1^{+0.7}_{-0.4}\times10^{12}~M_{\odot}$, which explains only $25-60$\% of $M_{\rm tot}$. We tested an NFW density profile instead of the $\beta$ model, 
and we obtained $M_{\rm E,~NFW}=1.1^{+1.2}_{-0.5}\times10^{12}~M_{\odot}$, which can explain $21-76$\% of $M_{\rm tot}$.
Note that, the center position of the main cluster has a significant offset from the {\it Eye of Horus}, and the galaxy distribution suggests that the {\it Eye of Horus} has complex mass structures along the line of sight. There might be substructures along the line of sight, which are not considered in this work.

\indent This is the first X-ray follow-up observation of the strong lens system discovered by Subaru-HSC, and we are planning to observe other strong lens systems with {\it XMM-Newton} and {\it Chandra}. Detailed modeling of the lensing of the {\it Eye of Horus} will be done, taking into account the result of this paper. When these observation and modeling are completed, we will be able to obtain robust constraint of the cosmological parameters and the gravitational structure of the distant galaxy.

\section*{Acknowledgements}
The Hyper Suprime-Cam (HSC) collaboration includes the astronomical communities of Japan and Taiwan, and Princeton University. The HSC instrumentation and software were developed by the National Astronomical Observatory of Japan (NAOJ), the Kavli Institute for the Physics and Mathematics of the Universe (Kavli IPMU), the University of Tokyo, the High Energy Accelerator Research Organization (KEK), the Academia Sinica Institute for Astronomy and Astrophysics in Taiwan (ASIAA), and Princeton University. Funding was contributed by the FIRST program from Japanese Cabinet Office, the Ministry of Education, Culture, Sports, Science and Technology (MEXT), the Japan Society for the Promotion of Science (JSPS), Japan Science and Technology Agency (JST), the Toray Science Foundation, NAOJ, Kavli IPMU, KEK, ASIAA, and Princeton University.

\indent This paper makes use of software developed for the Large Synoptic Survey Telescope. We thank the LSST Project for making their code available as free software at http://dm.lsst.org

\indent The Pan-STARRS1 Surveys (PS1) have been made possible through contributions of the Institute for Astronomy, the University of Hawaii, the Pan-STARRS Project Office, the Max-Planck Society and its participating institutes, the Max Planck Institute for Astronomy, Heidelberg and the Max Planck Institute for Extraterrestrial Physics, Garching, The Johns Hopkins University, Durham University, the University of Edinburgh, Queen ’ s University Belfast, the Harvard-Smithsonian Center for Astrophysics, the Las Cumbres Observatory Global Telescope Network Incorporated, the National Central University of Taiwan, the Space Telescope Science Institute, the National Aeronautics and Space Administration under Grant No. NNX08AR22G issued through the Planetary Science Division of the NASA Science Mission Directorate, the National Science Foundation under Grant No. AST-1238877, the University of Maryland, and Eotvos Lorand University (ELTE) and the Los Alamos National Laboratory.

\indent This work is supported in part by the Ministry of Science and Technology of Taiwan (grant MOST 106-2628-M-001-003-MY3) and by Academia Sinica (grant AS-IA-107-M01).

\indent This work is also supported in part by World Premier International Research Center Initiative (WPI Initiative), MEXT, Japan, and JSPS KAKENHI Grants Number JP15H05892, JP17H02868, JP18K03693 and JP19H05189.

\indent K.C.W. is supported in part by an EACOA Fellowship awarded by the East Asia Core Observatories Association, which consists of the Academia Sinica Institute of Astronomy and Astrophysics, the National Astronomical Observatory of Japan, the National Astronomical Observatories of the Chinese Academy of Sciences, and the Korea Astronomy and Space Science Institute.





\bibliographystyle{mnras}
\bibliography{EoH_paper} 








\bsp	
\label{lastpage}
\end{document}